\input lecproc.cmm
\input psfig.tex

\contribution{The terminal bulk Lorentz factor 
of relativistic electron-positron jet}
\author{Nicolas Renaud, Gilles Henri}
\contributionrunning{The terminal bulk Lorentz factor 
of relativistic $e^+-e^-$ jet}
\authorrunning{N. Renaud, G.Henri}
\address{Laboratoire d'Astrophysique, Observatoire de Grenoble, B.P 53X,
 F38041 Grenoble Cedex, France}
\abstract{We present numerical simulation of bulk Lorentz factor of 
relativistic electron-positron jet driven by Compton rocket effect from
 accretion disk radiation. The plasma is assumed to have a power-law 
distribution $n_{e}(\gamma)
\propto \gamma^{-s}$ ( $1 < \gamma
< \gamma_{max}$) 
and is continuously reheated
 to compensate radiation losses.
 We include Klein-Nishina (hereafter KN) corrections, and
 study the role of energy upper cut-off $\gamma_{max}$, spectral index $s$, and
 source compactness.
We determine terminal bulk Lorentz factor in the case of
 supermassive black holes relevant to AGN and stellar black holes relevant to 
galactic microquasars. In the latter case, effects of KN corrections 
are more important, and induce terminal bulk Lorentz factor $\gamma_{b\infty}$ 
smaller than
 in the former case. The result can explain the low bulk Lorentz factors for 
galactic sources (GRS1915+105, GROJ1655-40) compared to extragalactic ones. 
We also take into account the influence of the size of the accretion disk;
 if the external radius is small enough, the bulk Lorentz factor can be 
as high as 60, which is comparable to the values needed to explain 
extragalactic gamma-ray bursts.}

\titlea{1}{Introduction}
Superluminal motion observed in active galactic nuclei (AGN), 
especially in the class of blazars,
seems to be closely linked with 
high-energy emission. Such motion was recently observed in the Galaxy 
(Mirabel \& Rodriguez, 1994, Hjellming \& Rupen, 1995, Tingay, S. J., 
{\it et al.}, 1995) 
in the so-called microquasars. 
Nevertheless differences
are underlined in those two cases. The latter systems were observed with
small value of bulk Lorentz factor (around 2.5),
 while in the former ones values of
about 10-20 are frequent. 
A possible acceleration mechanism is the so-called 'Compton rocket' effect
 (O'Dell, 1981), {\it i.e.} anisotropic inverse Compton effect on a highly
 relativistic plasma.
It was then argued (Phinney
, 1982) that because of Compton cooling, only small value of $\gamma_{b\infty}$
 could be reached by this mechanism.
However, in the frame of the 'two-flow' model (Sol, Pelletier \& Ass\'eo, 
1989), a pair plasma can be reheated by the turbulence triggered by a
 surrounding jet (Henri \& Pelletier, 1991) and then the Compton rocket
 can be more efficient (Marcowith {\it et al.}, 1995). 
We here describe more precisely our 
assumptions and present our main results on acceleration of such 
pair plasma.

\titlea{2}{Compton rocket effect with Klein-Nishina corrections}

The pair plasma 
is assumed to be described by an energy distribution $n_e^\prime
(z,\gamma^\prime)
 \propto \gamma^{\prime -s}$ for $\gamma_{min} < \gamma^\prime
 < \gamma_{max}$, with
 $s$, $\gamma_{min}$ and $\gamma_{max}$ independant on $z$.
 The radiation force is due to soft photon coming from a standard 
accretion disk (Shakura \& Sunyaev, 1973) around a Schwarzschild black
 hole. 

Let $ {d \vec{p}^\prime \over dt^\prime}$ be the rate of momentum 
change for one electron and $\beta^\prime c=\sqrt{1-1/\gamma^{\prime 2}}c$ 
its velocity.
 The force exerced upon the pair plasma is then in the blob rest
 frame:

$$ F^{\prime z}=\int d\Omega_e^\prime d\gamma^\prime 
 n_e^\prime(\gamma^\prime,\Omega_e^\prime)  {d \vec{p}^\prime
 \over dt^\prime}.\vec{u_z} ,\eqno (1)$$
 
where $\vec{u_z}$ defines the jet axis direction. 
To determine $F^{\prime z}$, one can use two different approximations:
 for low energy particules, the KN cross-section tends to the Thomson
 limit and we can apply the result of Phinney (1982) for an isotropic
 pair plasma.
For high energy particules, KN corrections become important but we
 can assume that the head-on approximation is valid. 
 In this approximation we determine both momentum and energy loss rates 
in the electron rest frame
 and rely them to the blob rest frame using Lorentz 
transformation (see Renaud \& Henri, 1997, for detailed calculations.). 
We connect the two regimes by defining a critical Lorentz factor
$\gamma^\prime_{crit}$ for which errors in the head-on approximation
 ($\gamma^\prime >
 \gamma_{crit}$) and in
 the Thomson limit ($\gamma^\prime < \gamma_{crit}$) are of same order. 
This gives $\gamma_{crit} \sim
 (2 \langle \varepsilon \rangle)^{-1/3}$. $\langle \varepsilon \rangle 
\propto \dot{M}^{1/4} M^{-1/2}$, is the average photon energy emitted
 from accretion disk, $\dot{M}$ is the accretion rate and $M$ the mass of 
the central black hole.
For AGN $\langle \varepsilon \rangle \sim 10^{-4}$  and
 $\gamma_{crit}\sim 20-30$ while
 for a microquasar $\langle \varepsilon \rangle \sim 10^{-2}$ and
 $\gamma_{crit}\sim 5$.
Following Phinney (1982) we determine the acceleration of pair plasma by
considering the conservation of stress-energy tensor. 
For a reheated relativistic plasma 
(with $p^\prime=\rho^\prime / 3$) one finds the 
equation of motion.
$${d \gamma_b \over d z}={F^{\prime z} \over \rho^\prime} {1 \over {1
\over 3\gamma_b^2}+1}
, \eqno (2)$$

\titlea{3}{Results}
\titleb{3.1}{Effect of Klein-Nishina corrections}
The effect of the KN corrections
 is to reduce the contribution of high energetic collisions in
 the net radiation force seen by the plasma. This leads 
to smaller value of 
terminal Lorentz factor than expected in the Thomson regime.
 KN corrections begin to dominate roughly
 when $\gamma_{max}\langle\varepsilon \rangle \sim 1$.
The importance of this effect is so predominant for microquasars in
which accretion disk radiates more energetic photons than in the 
extragalactic case. 

\titleb{3.2}{Influence of the different parameters}
We determine terminal Lorentz factor for different configurations:
 we consider the case of supermassive black holes $M=10^9 M_{\odot}$
 relevant to AGN and stellar black holes $M=5M_{\odot}$ relevant to 
 microquasars.
For these two typical cases we show figure 1
 the influence of the spectral index and the energy cut-off $\gamma_{max}$.
\begfig -0.75 cm 
\psfig{rheight=7.5cm,width=12cm,height=8cm,angle=-90,file=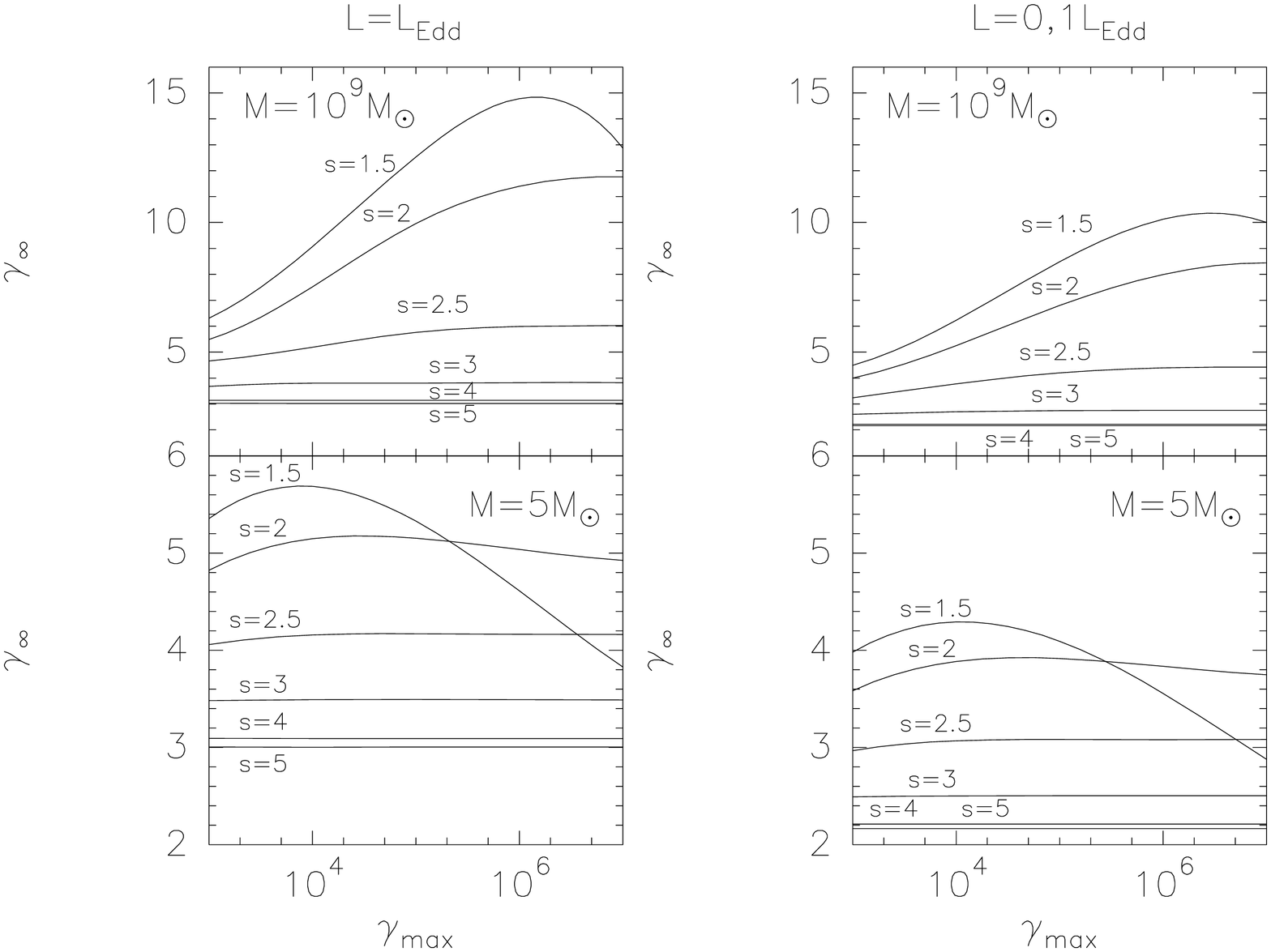}
\figure{1}{
Terminal Lorentz factor $\gamma_{b\infty}$ as a function
of $\gamma_{max}$ for different value of spectral index $s$. The two top
 pannels correspond to $M=10^9 M_{\odot}$ black hole with $r_e=3.10^3 r_g$,
 $L=L_{Edd}$ (left) and $L=0.1L_{Edd}$(right). The two bottom
 pannels correspond to $M=5 M_{\odot}$ black hole with $r_e=3.10^3 r_g$,
 $L=L_{Edd}$ (left) and $L=0.1L_{Edd}$(right).}
\endfig

\begfig -1.5 cm 
\psfig{rheight=4cm,width=12cm,height=8cm,angle=-90,clip=,file=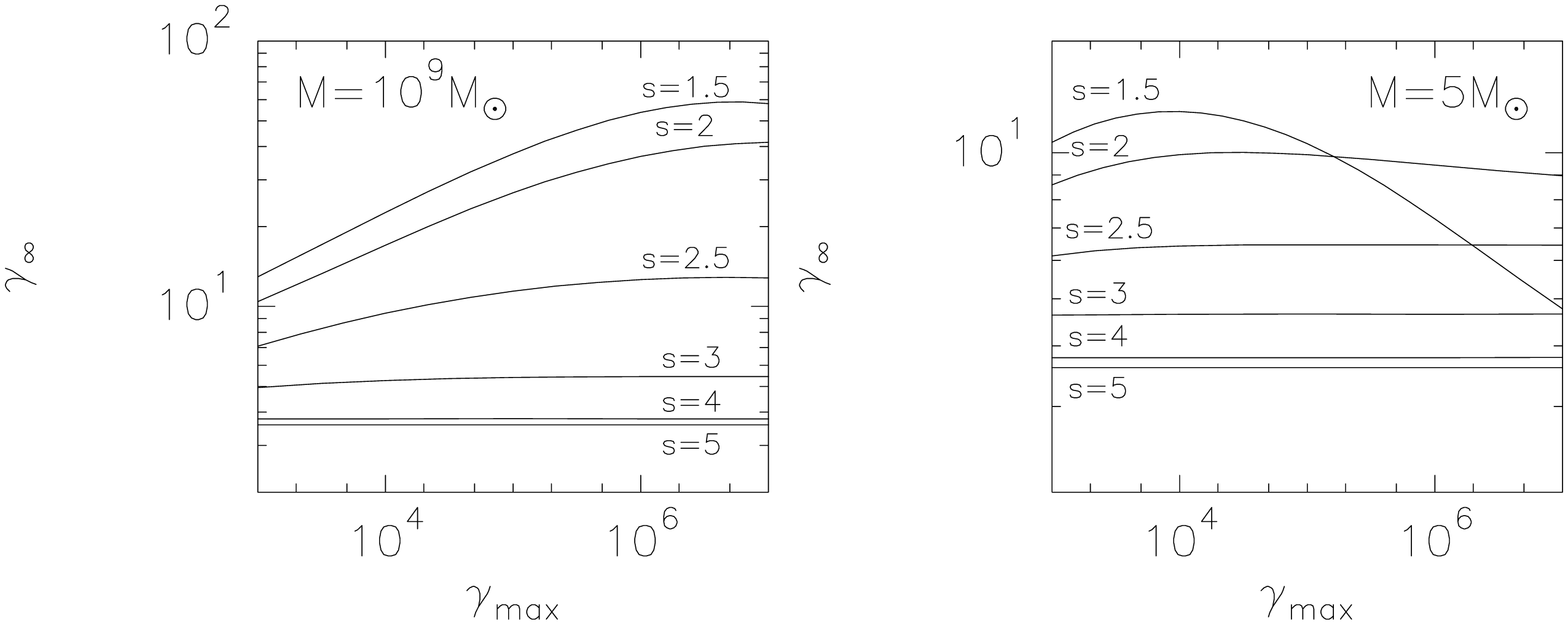}
\figure{2}{Terminal Lorentz factor $\gamma_{b\infty}$ as a function
of $\gamma_{max}$ for different value of spectral index $s$. The two
 pannels correspond to $r_e=10 r_g$ and $L=L_{Edd}$, 
$M=10^9 M_{\odot}$(left) $M=5 M_{\odot}$ (right).}
\endfig

For $ s < 3 $, $F^{\prime z}$ is dominated by the high
 value of $\gamma^\prime$. Terminal Lorentz factor admits then
a maximum
 when KN corrections begin to dominate. 
This value is more rapidily reached for a stellar black hole, explaining 
smaller $\gamma_{b \infty}$ than in extragalactic configurations.
 For $s > 3 $, $F^{\prime z}$ is dominated by the lower cut-off of the electronic distribution, explaining why no variation is evident
 with $\gamma_{max}$ in this case (see Renaud \& Henri, 1997). 
The plasma behaves as a cold one, and one 
finds small value of terminal Lorentz factor, like Phinney's result.
The dependance on $r_e$ is shown figure 2, where we extend the previous 
calculation to the case $r_e=10 r_G$. We find the same global 
behaviour of $\gamma_{b \infty}$ with spectral index and $\gamma_{max}$.
 In this configuration the radiation force is more efficient and 
 $\gamma_{b\infty}$ can be as high as 60 in the extragalactic case. 

\titlea{4}{Conclusion}
We find that for a given luminosity, the terminal Lorentz factor 
$\gamma_{b\infty}$
 admits a maximum due to KN corrections. They are predominant 
for stellar black holes leading to smaller values than 
for a supermassive 
black hole. Moreover the observed value of Lorentz factor of 
about 2.5 for the two
 microquasars with large spectral indexes (respectively 
$s \sim 4$, Finoguenov {\it et al.}, 1994 and $s \sim 4.6$, 
Harmon {\it et al.}, 1995) are coherent with our results.
 We also find that for fixed mass, spectral index, $\gamma_{b\infty}$ increases
 with decreasing $r_e$, reaching 60 for $r_e=10r_g$, 
which is comparable to the value needed
 to explain observed flux in gamma-ray burst. 
\bigskip\noindent
\begrefchapter{References}
\ref Finoguenov, A., et {\it al.}, 1994, ApJ, 424, 940
\ref Harmon, B. A., et {\it al.}, 1995, Nature, 374, 703
\ref Henri, G., \& Pelletier, G., 1991, ApJ Letters, 383, L7
\ref Hjellming, R. M., \& Rupen, M. P., 1995, Nature, 375, 464
\ref Marcowith, A., Henri, G., \& Pelletier, G., 1995, MNRAS, 277, 681
\ref Mirabel, I. F., \& Rodriguez, L. F., 1994, Nature, 371, 46
\ref O'Dell, S. L., 1981, ApJ Letters, 243, L147
\ref Phinney, E. S., 1982, MNRAS, 198, 1109
\ref Renaud, N. \& Henri, G., 1997 , MNRAS, in preparation
\ref Shakura, N. I., \& Sunyaev, R. A. , 1973, A\&A, 24, 337
\ref Sol, H., Pelletier, G., Ass\'eo, E., 1989, MNRAS, 327, 411
\ref Tingay, S. J., et {\it al.}, 1995, Nature, 374, 141
\endref

\byebye
